\documentclass[12pt]{article}
\usepackage{a4wide}
\usepackage{epsfig}
\usepackage{amsmath}

\newcommand{\half}{{\textstyle\frac{1}{2}}}

\newlength{\absize}
\catcode`@=11
\def\citer{\@ifnextchar [{\@tempswatrue\@citexr}{\@tempswafalse\@citexr[]}}

\def\@citexr[#1]#2{\if@filesw\immediate
  \write\@auxout{\string\citation{#2}}\fi
  \def\@citea{}\@cite{\@for\@citeb:=#2\do
    {\@citea\def\@citea{--\penalty\@m}\@ifundefined
       {b@\@citeb}{{\bf ?}\@warning
       {Citation `\@citeb' on page \thepage \space undefined}}%
\hbox{\csname b@\@citeb\endcsname}}}{#1}}
\catcode`@=12

\begin{document}
  \thispagestyle{empty}
  \pagestyle{empty}
  \renewcommand{\thefootnote}{\fnsymbol{footnote}}
\newpage\normalsize
    \pagestyle{plain}
    \setlength{\baselineskip}{4ex}\par
    \setcounter{footnote}{0}
    \renewcommand{\thefootnote}{\arabic{footnote}}
\newcommand{\preprint}[1]{%
  \begin{flushright}
    \setlength{\baselineskip}{3ex} #1
  \end{flushright}}
\renewcommand{\title}[1]{%
  \begin{center}
    \LARGE #1
  \end{center}\par}
\renewcommand{\author}[1]{%
  \vspace{2ex}
  {\Large
   \begin{center}
     \setlength{\baselineskip}{3ex} #1 \par
   \end{center}}}
\renewcommand{\thanks}[1]{\footnote{#1}}
\begin{flushright}
\end{flushright}
\vskip 0.5cm

\begin{center}
{\large \bf  A $q$-deformed Uncertainty Relation}
\end{center}
\vspace{1cm}
\begin{center}
Jian-zu Zhang$^{a,b,*}$
\end{center}
\vspace{1cm}
\begin{center}
$^a$ Institut f\"ur Theoretische Physik, RWTH-Aachen,
D-52056 Aachen, Germany\\
$^b$ Institute for Theoretical Physics, Box 316,
East China University of Science and Technology,
Shanghai 200237, P. R. China
\end{center}
\vspace{1cm}

\begin{abstract}
Within the formulation of a $q$-deformed Quantum Mechanics a
qualitative undercut of the $q$-deformed uncertainty relation from
the Heisenberg uncertainty relation is revealed. When $q$ is some
fixed value not equal to one, recovering of ordinary quantum
mechanics and the corresponding recovering condition are
discussed.
\end{abstract}

\begin{flushleft}
${^*}$ E-mail address: jzzhangw@ecust.edu.cn  \\
\hspace{3.1cm}  jzzhang@physik.uni-kl.de
\end{flushleft}
\clearpage
The Heisenberg uncertainty relation is a direct result of the
Heisenberg commutation relation (Heisenberg algebra). According to
the present tests of quantum electrodynamics quantum theories
based on the Heisenberg algebra are correct at least down to
$10^{-17}$ cm. A question arises whether there is a possible
modification of the Heisenberg algebra at short distances much
smaller than $10^{-17}$ cm.
 In search for such possibilities at short distances (or high
 energy scales) consideration of the space structure is a useful
guide. Recently as a possible candidate of short distance new
physics a $q$-deformed quantum mechanics is proposed
\citer{Fichtmuller,Kempf} in the framework quantum group. Quantum
groups are a generalization of symmetry groups which have been
successfully used in physics. A general feature of spaces carrying
a quantum group structure is that they are noncommutative and
inherit a well-defined mathematical structure from the quantum
group symmetries. In applications in physics questions arise
whether the structure can be used for physics at short distances
and what phenomena could be linked to it. Starting from such a
noncommutative space as configuration space a generalisation to a
phase space is obtained \cite{Fichtmuller}. Such noncommutative
phase space is a $q$-deformation of the quantum mechanical phase
space and thus all the machinery used in quantum mechanics can be
applied in $q$-deformed quantum mechanics \citer{Fichtmuller,LRW}.
A $q$-deformed Heisenberg algebra, as a generalization of
Heisenberg,s algebra, is established \cite{Fichtmuller,Hebecker}
in $q$-deformation phase space.

Starting from the $q$-deformed Heisenberg algebra Ref.~\cite{JZZ}
obtained a $q$-deformed uncertainty relation and found that the
Heisenberg uncertainty relation is undercut. This is a qualitative
deviation from the Heisenberg uncertainty relation. It therefore
raises the question when $q$ is some fixed value not equal to one
where or not the Heisenberg uncertainty relation is recovered and
what is the corresponding condition. In this letter we investigate
the above important open question in the framework of the
$q$-deformed harmonic oscillator proposed in Ref.~\cite{LRW}. We
find a further qualitative deviation of the $q$-deformed
uncertainty relation from the Heisenberg uncertainty relation.

The $q$-deformed harmonic oscillators were first studied by
Macfarlane \cite{Macfarlane} and Biedenharn \cite{Biedenharn}.
Ref.~\cite{LRW} find a general ansatz of the creation and
annihilation operators in terms of the $q$-deformed phase space
variables, the position operator $X,$ momentum operator $P,$ and
scaling operator $U$ which satisfy the $q$-deformed Heisenberg
algebra
\begin{equation}
\label{Eq:q-algebra} q^{1/2}XP-q^{-1/2}PX=i\hbar U, \qquad
UX=q^{-1}XU, \qquad UP=qPU,
\end{equation}
where $X$ and $P$ are hermitian and $U$ is unitary:
\begin{equation}
\label{Eq:hermitian} X^{\dagger}=X, \qquad P^{\dagger}=P, \qquad
U^{\dagger}=U^{-1}.
\end{equation}
In (\ref{Eq:q-algebra}) the parameter $q$ is real and $q>1$. The
operator $U$ closely relates to properties of dynamics and plays
an essential role in $q$-deformed quantum mechanics. The
definition of the algebra (\ref{Eq:q-algebra}) is based on the
definition of the hermitian momentum operator $P$. However, if $X$
is assumed to be a hermitian operator in a Hilbert space the usual
quantization rule $P\to -i\partial_X$ does not yield a hermitian
momentum operator. Ref.~\cite{Fichtmuller} showed that a hermitian
momentum operator $P$ is related to $\partial_X$ and $X$ in a
nonlinear way by introducing a scaling operator $U$
\begin{eqnarray}
U^{-1}\equiv q^{1/2}[1+(q-1)X\partial_X], \qquad
\bar\partial_X\equiv -q^{-1/2}U\partial_X, \qquad P\equiv
-\frac{i}{2}(\partial_X-\bar\partial_X), \nonumber
\end{eqnarray}
Where $\bar\partial_X$ is the conjugate of $\partial_X$. From
(\ref{Eq:q-algebra})and (\ref{Eq:hermitian}) it follows that the
$q$-deformed commutation relation is
\begin{equation}
\label{Eq:[X,Y]o} XP - PX = i\hbar\frac{U + U^{-1}}{q^{1/2} +
q^{-1/2}},
\end{equation}
which yields a $q$-deformed uncertainty relation and shows an
undercut of Heisenberg's minimal uncertainty relation. Because of
the complicated relations among $X,$ $P$ and $U,$ from the above
equation it is not clear when $q$ is some fixed value not equal to
one whether Heisenberg's uncertainty relation can be recovered. We
now investigate this question in an equavalent framework of
algebra (\ref{Eq:q-algebra}), the $q$-deformed harmonic oscillator
proposed in Ref.~\cite{LRW}.

The expression for annihilation and  creation operators $a$ and
$a^\dagger$ in terms of $X,$ $P$, and $U$ are
\begin{equation}
\label{Eq:Wess-Ansatz} a=\alpha U^{-2M}+\beta U^{-M}P, \qquad
a^\dagger=\bar\alpha U^{2M}+\bar\beta PU^{M},
\end{equation}
where $M=0,1,2,\ldots$, $\alpha$ and $\beta$ are complex numbers.
From the ansatz (\ref{Eq:Wess-Ansatz}) it follows that $a$ and
$a^\dagger$ satisfy the following algebra:
\begin{equation}
\label{Eq:a-adagger} a a^\dagger -q^{-2M} a^\dagger a=1
\end{equation}
with the condition (up to a phase of $\alpha$)
\begin{equation}
\label{Eq:phase} \alpha=\frac{e^{i\phi}}{(1-q^{-2M})^{1/2}}.
\end{equation}
The ansatz (\ref{Eq:Wess-Ansatz}) is determined by the requirement
of the equivalence of algebras (\ref{Eq:q-algebra}) and
(\ref{Eq:a-adagger}). The operators $a$ and $a^\dagger$ are
related to the operator $X$ in a complicated way (In
(\ref{Eq:Wess-Ansatz}) $X$ is nonlinearly included in the operator
$U$.)

The $q$-deformed phase space variables $X$, $P$ and the scaling
operator $U$ can be expressed in terms of the usual canonical
variables $\hat x$ and $\hat p$ as follows \cite{Fichtmuller}:
\begin{equation}
\label{Eq:X-x} X=\frac{[\hat z+\half]}{\hat z+\half}\hat x, \quad
P=\hat p, \quad U=q^{\hat z},
\end{equation}
where $\hat z=-i(\hat x\hat p+\hat p\hat x)/2\hbar$,
$[A]=(q^A-q^{-A})/(q-q^{-1})$, and $\hat x$ and $\hat p$ satisfy
\begin{equation}
\label{Eq:x-p} [\hat x,\hat p]=i\hbar,\quad  \hat x^\dagger=\hat
x, \quad \hat p^\dagger =\hat p.
\end{equation}
From (\ref{Eq:X-x}) and (\ref{Eq:x-p}) it follows that $X$, $P$
and $U$ satisfy (\ref{Eq:q-algebra}) and (\ref{Eq:hermitian}). The
algebra (\ref{Eq:x-p}) is realized as follows:
\begin{equation}
\label{Eq:x-x} \hat x=x, \quad \hat
p=p+\frac{\gamma}{\sqrt{1-q^{-2M}}}, \quad p=-i\hbar\partial_x,
\end{equation}
where $\gamma$ is a real constant. Let $q=e^f$, ($0<f\ll1$). In
the limit $f\to 0,$ there are two singular terms in the expression
of $a$ in (\ref{Eq:Wess-Ansatz}). The condition of cancellation of
two singular terms is
\begin{equation}
\label{Eq:beta-gamma} \beta\gamma=-e^{i\phi},
\end{equation}
which leads to $\alpha\bar\beta=\bar\alpha\beta.$ In the limit
$f\to 0,$ from (\ref{Eq:Wess-Ansatz}), (\ref{Eq:phase}),
(\ref{Eq:X-x}), (\ref{Eq:x-x}) and (\ref{Eq:beta-gamma}) it
follows that
\begin{equation}
\label{Eq:a-a0} a \to e^{i\phi}\left(\frac{1}{2}i\gamma
x-\frac{1}{\gamma}p\right)=a_0.
\end{equation}
The usual expression
$a_0=\sqrt{m\omega/2\hbar}x+i\sqrt{1/2m\omega\hbar}p$ is recovered
if we choose
\begin{equation}
\label{Eq:phi} e^{i\phi}=\mp i,\quad
\gamma=\pm\sqrt{2m\omega},\quad \beta=\frac{i}{\sqrt{2m\omega}}.
\end{equation}

The $q$-deformed Hamiltonian is
\begin{equation}
\label{Eq:H} H_{\omega}=\hbar\omega a^\dagger a.
\end{equation}
In the limit $f\to 0,$ (\ref{Eq:H})  reduces to the undeformed one
$H_{\omega}\to H_0=\hbar\omega a_0^\dagger a_0.$

In order to reveal the possible deviation of the $q$-deformed
uncertainty relation yielded by the algebra (\ref{Eq:q-algebra}),
or the equivalent algebra (\ref{Eq:a-adagger}), from Heisenberg's
uncertainty relation, we decompose $a$ and $a^\dagger$ into a pair
of quadrature operators $Q$ and $K$
\begin{equation}
\label{Eq:Q-K} a=\frac{Q}{2D_1}+i\frac{K}{2D_2},\quad
a^\dagger=\frac{Q}{2D_1}-i\frac{K}{2D_2}.
\end{equation}
Where $Q$ and $K$ are hermitian operators, $D_1$ and $D_2$ are
positive real parameters and have, respectively, the dimensions of
the position and momentum. Introducing (\ref{Eq:Q-K}) into
(\ref{Eq:a-adagger}), we obtain an equivalent $q$-deformed
commutation relation of the algebra (\ref{Eq:q-algebra}):
\begin{equation}
\label{Eq:[Q,K]}
QK-KQ=i\hbar-i\hbar(1-q^{-2M})\left(\frac{Q^2}{4D_1^2}
+\frac{K^2}{4D_2^2}\right)
\end{equation}
with a condition
\begin{equation}
\label{Eq:DD} \frac{4D_1 D_2}{1+q^{-2M}}=\hbar.
\end{equation}

From (\ref{Eq:[Q,K]}) we now study minimal uncertainties in the
position and momentum. We start with \cite{Kempf},
\begin{equation}
\label{Eq:uncer-1} |[(Q-\bar Q)+i\eta(K-\bar K)]|i\rangle|^2\ge 0,
\end{equation}
where $\bar F=\langle i|F|i\rangle.$ For any real $\eta,$
(\ref{Eq:[Q,K]}) and (\ref{Eq:uncer-1}) yield
\begin{equation}
\label{Eq:uncer-2} (\Delta K)^2\left[\eta-\frac{\hbar A}{2(\Delta
K)^2}\right]^2-\frac{\hbar^2 A^2}{4(\Delta K)^2}+(\Delta Q)^2\ge
0,
\end{equation}
\begin{equation}
\label{Eq:A} A=1-(1-q^{-2M})\left[\frac{(\Delta Q)^2+(\bar
Q)^2}{4D_1^2}+\frac{(\Delta K)^2+(\bar K)^2}{4D_2^2}\right],
\end{equation}
where $\Delta F=\sqrt{\langle i|(F-\bar F)^2|i\rangle}.$ Choosing
$\eta=\hbar A/2(\Delta K)^2,$  (\ref{Eq:uncer-2}) yields the
following uncertainty relation:
\begin{equation}
\label{Eq:Delta-Q,K} \Delta Q \Delta K
\ge\frac{\hbar}{2}-\frac{\hbar}{2}(1-q^{-2M})\left[\frac{(\Delta
Q)^2+(\bar Q)^2}{4D_1^2} +\frac{(\Delta K)^2+(\bar
K)^2}{4D_2^2}\right].
\end{equation}
Because of $1-q^{-2M}>0,$ (\ref{Eq:Delta-Q,K}) shows that the
Heisenberg minimal uncertainty relation $\Delta Q \Delta K
=\frac{\hbar}{2}$ can be undercut \cite{JZZ}.

Defining
\begin{equation}
\label{Eq:f-Delta-Q,K} f(\Delta Q, \Delta K) =\Delta Q \Delta
K-\frac{\hbar}{2}\left\{1-(1-q^{-2M})\left[\frac{(\Delta
Q)^2+(\bar Q)^2}{4D_1^2} +\frac{(\Delta K)^2+(\bar
K)^2}{4D_2^2}\right]\right\},
\end{equation}
conditions of $(\Delta K)_{min}$ (or $(\Delta Q)_{min}$)
\begin{eqnarray}
\frac{\partial}{\partial {\Delta Q}}f(\Delta Q, \Delta K)=0 \quad
\left(or \quad \frac{\partial}{\partial {\Delta K}}f(\Delta Q,
\Delta K)=0\right),\quad f(\Delta Q, \Delta K) =0 \nonumber
\end{eqnarray}
yield
\begin{equation}
\label{Eq:condition-1}
\left[\frac{\hbar(1-q^{-2M})}{4D_1^2}\right]\Delta Q+\Delta K=0
\end{equation}
or
\begin{equation}
\label{Eq:condition-2} \Delta
Q+\left[\frac{\hbar(1-q^{-2M})}{4D_2^2}\right]\Delta K=0.
\end{equation}
(\ref{Eq:condition-1}) (or (\ref{Eq:condition-2})) shows that the
only non-negative solution is
\begin{equation}
\label{Eq:mini} (\Delta Q)_{min}=(\Delta K)_{min}=0
\end{equation}
and $f((\Delta Q)_{min},(\Delta K)_{min})=0$ yields
\begin{equation}
\label{Eq:condition-3} \frac{(\bar Q)^2}{D_1^2} +\frac{(\bar
K)^2}{D_2^2}=\frac{4}{1-q^{-2M}}.
\end{equation}

In the limit $q\to 1,$ from (\ref{Eq:a-a0}), (\ref{Eq:phi}),
(\ref{Eq:Q-K}) and (\ref{Eq:DD}) it follows that
\begin{equation}
\label{Eq:condition-4} D_1=d_1(q)\sqrt{\hbar/2m\omega},\quad
D_2=d_2(q)\sqrt{m\omega\hbar/2}, \quad
d_1(q)d_2(q)=\frac{1}{2}(1+q^{-2M}),
\end{equation}
where $d_1$ and $d_2$ are dimensionless and satisfy the limiting
conditions
\begin{equation}
\label{Eq:condition-5} d_1(q)_{q\to 1}\to 1, \quad d_2(q)_{q\to
1}\to 1.
\end{equation}

When $q\to 1,$ Eq.~(\ref{Eq:Delta-Q,K}) reduces to the Heisenberg
uncertainty relation. But when $q$ is some fixed value not equal
to one, (\ref{Eq:mini}) shows that there are states when the
condition (\ref{Eq:condition-3}) is met $\Delta Q$ and $\Delta K $
can simultaneously equal to zero. This is a qualitative deviation
from the Heisenberg uncertainty relation. It therefore raises a
question whether or not the Heisenberg uncertainty relation is
recovered for fixed $q$ and what is the recovering condition. In
fact, (\ref{Eq:[Q,K]}) shows that when $||Q^2||$ and $||K^2||$
satisfy ($||A||$ is the norm of $A$)
\begin{equation}
\label{Eq:condition-5} 0\le ||Q^2||\le \frac{4D_1^2}{1-q^{-2M}},
\quad 0\le ||K^2||\le \frac{4D_2^2}{1-q^{-2M}}
\end{equation}
the ordinary quantum mechanical behavior is approximately
reproduced.

The canonical conjugate pair of operators $Q$ and $K$ defined in
(\ref{Eq:Q-K}) are not the position and momentum. But from
(\ref{Eq:a-a0}), (\ref{Eq:phi}), (\ref{Eq:Q-K}),
(\ref{Eq:condition-4}) and (\ref{Eq:condition-5}) it follows that
in the limit $q\to 1,$ $Q$ and $K$ approach, respectively, the
undeformed position operator $x$ and momentum operator $p.$

The correct Hamiltonian of the $q$-deformed harmonic oscillators
is $H_{\omega}$ defined in (\ref{Eq:H}). But the Hamiltonian
$H_{Q,K}=\frac{1}{2m}K^2+\frac{1}{2}m\omega^2 Q^2$ which is
extensively considered in literature is not the correct one,
though in the limit $q\to 1$ it approaches to
$H_{\omega}+\frac{1}{2}.$ Ref.~\cite{Nelson} also noticed the
differences between $H_{\omega}$ and $H_{Q,K}$ (Ref.~\cite{Nelson}
simply took $d_1(q)=d_2(q)=1$) that $H_{\omega}$  does possess
conventional physical properties but $H_{Q,K}$ probably does not
permit a consistent physical interpretation.

In order to further confirm the existence of the qualitative
deviation from the Heisenberg uncertainty relation, we investigate
the coherent states of electromagnetic fields by a different
method. For a single mode field of frequency $\omega$ the electric
field operator $E(t)$ is represented as $E(t)=E_0[\hat
a\exp(-i\omega t)+\hat a^\dagger\exp(i\omega t)]$ where $\hat a$
and $\hat a^\dagger$ are the photon annihilation and creation
operators. $\hat a$ and $\hat a^\dagger$ can be decomposed into a
pair of dimensionless conjugate quadrature operators $X_e$ and
$Y_e:$
\begin{equation}
\label{Eq:X,Y} \hat a=\frac{1}{2}(X_e+iY_e), \quad \hat
a^\dagger=\frac{1}{2}(X_e-iY_e)
\end{equation}
where $X_e$ and $Y_e$ are hermitean. In terms of $X_e$ and $Y_e$
the operator $E(t)$ is then expressed as $E(t)=E_0(X_e cos\omega
t+Y_e sin\omega t).$

Suppose $\hat a$ and $\hat a^\dagger$ satisfy the algebra
(\ref{Eq:a-adagger}). From  (\ref{Eq:condition-5}) and
(\ref{Eq:a-adagger}) it follows that
\begin{equation}
\label{Eq:[X,Y]} X_e Y_e-Y_e X_e=iC
\end{equation}
with
\begin{equation}
\label{Eq:C1} C=2-2(1-q^{-2M})\hat a^\dagger\hat a.
\end{equation}

We now prove that for any state
\begin{equation}
\label{Eq:C2} 0\le \langle C\rangle\le 2.
\end{equation}

Suppose the existence of a ground $|0\rangle$ satisfying $\hat
a|0\rangle=0$ and $\langle 0|0\rangle =1.$ Using the algebra
(\ref{Eq:a-adagger}) we obtain \cite{LRW}:
$|n\rangle=\epsilon^{-1/2}_n(\hat a^\dagger)^n|0\rangle,$ and
$\hat a^\dagger \hat a|n\rangle=\epsilon_n|n\rangle,$ $(n=0, 1, 2,
\cdots)$ with $\epsilon_n=(1-q^{-2nM})/(1-q^{-2M}).$ We notice
that in the limit $q\to 1,$ we have $\epsilon_n=n.$ Because $\hat
a^\dagger\hat a$ is hermitean, we may suppose that its eigen
states are complete, and any states $|\;\rangle$ may be expanded
as $|\;\rangle=\sum_{n=0}^\infty C_n|n\rangle$ with
$\sum_{n=0}^\infty |C_n|^2=1.$ Thus in any states we have
\begin{eqnarray}
\langle \hat a^\dagger\hat a
\rangle={\displaystyle\sum_{n=0}^\infty|C_n|^2\epsilon_n}
=(1-q^{-2M})^{-1}\left(1-
{\displaystyle\sum_{n=0}^\infty|C_n|^2q^{-2nM}}\right) \nonumber
\end{eqnarray}
and $\langle C\rangle=2\sum_{n=0}^\infty |C_n|^2q^{-2nM}.$ Thus
$0\le \langle C\rangle\le 2\sum_{n=0}^\infty |C_n|^2=2.$ From
(\ref{Eq:[X,Y]}) to (\ref{Eq:C2}) we conclude that Heisenberg's
minimal uncertainty relation $\Delta X_e \Delta Y_e=1$ is
undercut. We now investigate the $q$-deformed coherent states
which were investigated in Refs.~\cite{Codriansky,Solomon}. A
coherent state $|\beta\rangle$ satisfies $\hat a
|\beta\rangle=\beta|\beta\rangle.$ From (\ref{Eq:a-adagger}) and
(\ref{Eq:X,Y}) it follows that
\begin{equation}
\label{Eq:Delta-X,Y} (\Delta X_e)^2=(\Delta
Y_e)^2=1-(1-q^{-2M})|\beta|^2\ge 0.
\end{equation}
A similar result was obtained in Ref.~\cite{McDermott}. Specially,
we notice that when
\begin{equation}
\label{Eq:beta} |\beta|^2\to \frac{1}{1-q^{-2M}}
\end{equation}
(\ref{Eq:Delta-X,Y}) yield
\begin{equation}
\label{Eq:Delta-Xm,Ym} (\Delta X_e)_{min}=(\Delta Y_e)_{min}\to 0.
\end{equation}
For the coherent state $|\beta\rangle,$ from (\ref{Eq:X,Y}) it
yields $(\bar X_e)^2+(\bar Y_e)^2=4|\beta|^2.$ Thus the condition
(\ref{Eq:beta}) is just the condition (\ref{Eq:condition-3}).

(\ref{Eq:Delta-Q,K})  and (\ref{Eq:[X,Y]})-(\ref{Eq:C2}) show that
within the formulation of the algebra (\ref{Eq:q-algebra}) or
equalent algebra (\ref{Eq:a-adagger}) the Heisenberg uncertainty
relation is undercut. Specially, (\ref{Eq:mini}) and
(\ref{Eq:Delta-Xm,Ym}) show a qualitative deviation from the
Heisenberg uncertainty relation, that is, there are some special
states, which permit simultaneous zero minimal uncertainties in a
pair of conjugate variables. In a sense (\ref{Eq:Delta-Q,K}) may
be called '{\it semi-uncertainty relation}'. An example is
$q$-deformed coherent states of electromagnetic fields. In
(\ref{Eq:Delta-X,Y}) $(\Delta X_e)^2,$ $(\Delta Y_e)^2$ and
$|\beta|^2$ are quantities which can be measured by experiments.
In principle, (\ref{Eq:Delta-X,Y}) could test the deviation from
the Heisenberg uncertainty relation, and when the condition
(\ref{Eq:beta}) is met it should yield simultaneous zero
uncertainties in two quadratures of electric fields. Of course,
any attempt to test the possible $q$-deformed effects is
challenge, because if $q$-deformed quantum mechanics is a correct
theory at short distances, its corrections to the present-day
physics must be extremely small.

We conclude this paper by clarifying the following interesting
questions.

(I) Why the 'undercutting' occurs? In order to track down the
origin of this undercutting phenomena a deep understanding of the
$q$-deformed commutation relation (\ref{Eq:[X,Y]o}) which is the
manifestation of the non-commutativity structure is necessary. We
demonstrate that the expectation value of the operator $C_q=(U +
U^{-1})/(q^{1/2} + q^{-1/2})$ in (\ref{Eq:[X,Y]o}) satisfies
$\langle |C_q|\rangle\le 1.$

We notice that $U+U^{-1}$ is hermitean, but $U-U^{-1}$ is
anti-hermitean, for any state $\langle (U+U^{-1})^2\rangle\ge 0,
\langle (U-U^{-1})^2\rangle\le 0.$ Thus $\langle
(U+U^{-1})^2\rangle-4 =\langle (U-U^{-1})^2\rangle\le 0$, and
$(\langle U+U^{-1}\rangle)^2\le \langle (U+U^{-1})^2\rangle\le 4$,
i.e. $|\langle U+U^{-1}\rangle|\le 2$. Because
$q^{1/2}+q^{-1/2}\ge 2$ for any $q>0$, we obtain $|\langle
C_q\rangle|=|\langle U+U^{-1}\rangle|/(q^{1/2}+q^{-1/2})\le 1.$
Eq.~(\ref{Eq:[X,Y]o}) gives a $q$-deformed uncertainty relation
$\Delta X\Delta P\ge \hbar|\langle C_q\rangle|/2$ which shows that
the Heisenberg minimal uncertainty relation $\Delta X\Delta P=
\hbar/2$ can be undercut.

(II) Regarding (\ref{Eq:Delta-Q,K}) and (\ref{Eq:Delta-Xm,Ym}),
what are the number states $|n\rangle$ which in the expansion of
the coherent states $|\beta\rangle$  or the latter, or the states
$|i\rangle$ for the former, give the undercutting? In order to
answer this question, for the number states $|n\rangle$ we
calculate $(\Delta X_e)_n^2=(\Delta
Y_e)_n^2=1/4+(1+q^{-2M})\epsilon_n/4$ which shows that the number
states do not give the undercutting. When $q\to 1,$ we have
$\epsilon_n\to n,$  and $(\Delta X_e)_n^2$ and $(\Delta Y_e)_n^2$
reduce to the undeformed ones.

(III) What representation of $q$-oscillator algebras lead to the
undercutting of Heisenberg's uncertainty relation or to the
opposite case? In order to clarify this matter we investigate
$q$-deformed algebras considered in Ref.~\cite{Nelson} $a
a^\dagger -q^{\pm 1/2} a^\dagger a=q^{\mp N/2},$ where
$N=a^\dagger a$ is the number operator. These algebras involve $N$
explicitly, and thus are not natural realization of them. Defining
\cite{Macfarlane} $b_1=a q^{-(N-1)/4},$ $b^\dagger_1=
q^{-(N-1)/4}a^\dagger$  and $b_2=a q^{(N-1)/4},$ $b^\dagger_2=
q^{(N-1)/4}a^\dagger,$ respectively, for the algebra with
$q^{-N/2}$ and $q^{N/2},$ the above algebras with $q^{-N/2}$ and
$q^{N/2}$ reduce, respectively, to $b_1 b^\dagger_1 -q^{-1}
b^\dagger_1 b_1=1$ and $b_2 b^\dagger_2 -q b^\dagger_2 b_2=1.$ The
former is just the algebra (\ref{Eq:a-adagger}) with $M=1$ which
leads to undercutting of Heisenberg's uncertainty relation. The
later, introducing operators $X_2$ and $Y_2$ by
$b_2=\frac{1}{2}(X_2+iY_2), \quad
b^\dagger_2=\frac{1}{2}(X_2-iY_2)$, reduces to $X_2 Y_2-Y_2
X_2=iC_2$ with $C_2=2+2(q-1)b_2^\dagger b_2.$ By a similar
procedure of proving (\ref{Eq:C2}) we can prove that for any
states $\langle C_2\rangle\ge 2$ (The equal sign holds only for
the case $q=1$.) Thus the later leads to the opposite case.
\vspace{0.4cm}

{\bf Acknowledgement} The author would like to thank Prof. J. Wess
very much
 for his many stimulating helpful discussions, to thank Prof. H.
 A. Kastrup for many helpful comments. He would also like to thank
 Institut f\"ur  Theoretische Physik, RWTH-Aachen, for warm
 hospitality. His work has been supported by the Deutsche
 Forschungsgemeinschaft (Germany), the National Natural Science
Foundation of China under the Grant No. 19674014, and the Shanghai
Education Development Foundation.



\vspace{0.4cm}

\begin{thebibliography}{09}
\bibitem{Fichtmuller}   
M.~Fichtm\"uller, A.~Lorek and J.~Wess,
Z.\ Phys.\ {\bf C71} (1996) 533 [hep-th/9511106].

\bibitem{Schwenk} 
J.~Schwenk and J.~Wess,
Phys.\ Lett.\ {\bf B291} (1992) 273.

\bibitem{LW} 
A.~Lorek and J.~Wess,
Z.\ Phys.\ {\bf C67} (1495) 671 [q-alg/9502007].

\bibitem{LRW} 
A.~Lorek, A.~Ruffing and J.~Wess,
Z.\ Phys.\ {\bf C74} (1997) 369 [hep-th/9605161].

\bibitem{Hebecker} 
A.~Hebecker, S.~Schreckenberg, J.~Schwenk, W.~Weich and J.~Wess,
Z.\ Phys.\ {\bf C64} (1994) 355.

\bibitem{WZ} 
J. Wess and B. Zumino, Nucl.\ Phys.\ Proc.\ Suppl.\ {\bf 18B}
(1991) 302.

\bibitem{JZZ} 
Jian-zu Zhang, Phys.\ Lett.\  {\bf B440} (1998) 66.

\bibitem{Kempf} 
A.~Kempf,
J.\ Math.\ Phys.\ {\bf 35} (1994) 4483 [hep-th/9311147].

\bibitem{Macfarlane} 
A.~J.~Macfarlane,
J.\ Phys.\ {\bf A22} (1989) 4581.

\bibitem{Biedenharn} 
L.~C.~Biedenharn,
J.\ Phys.\ {\bf A22} (1989) L873.

\bibitem{Nelson} 
C.A. Nelson, M.H. Fields, Phys.\ Rev.\  {\bf A51} (1995) 2410.

\bibitem{Codriansky} 
S. Codriansky, Phys.\ Lett.\ {\bf A184} (1993) 381.

\bibitem{Solomon} 
A. I. Solomon, J. Katriel, J. Phys. {\bf A23} (1990) L1209; J.
Katriel, A. I. Solomon, J. Phys. {\bf A24} (1991) 2093.

\bibitem{McDermott} 
R.J. McDermott and A.I. Solomon, J. \ Phys.\  {\bf A 27} (1994)
L15.

\end{thebibliography}
\end{document}